\documentclass[twocolumn, aps, prl]{revtex4-2}

\usepackage{graphicx}
\usepackage{natbib}

\begin{document}

\title{Modulation doping of the FeSe monolayer on SrTiO$_3$}

\author{Fengmiao Li, Ilya Elfimov and George A. Sawatzky}

\affiliation{Department of Physics \& Astronomy, University of British Columbia, Vancouver, British Columbia, Canada V6T 1Z1}
\affiliation{Quantum Matter Institute, University of British Columbia, Vancouver, British Columbia, Canada V6T 1Z4}

\date{\today}

\begin{abstract}
The discovery of higher-temperature superconductivity in FeSe monolayers on SrTiO$_3$ (STO) substrates has sparked a surge of interest in the interface superconductivity. One point of the agreement reached to date is that modulation doping by impurities in the substrate is critical for the enhanced superconductivity. Remarkably, the universal doping of about 0.1 electrons per Fe, \textit{i.e.}, so-called ``magic'' doping, has been observed on a range of Ti oxide substrates, which concludes that there likely is some important interaction limiting the FeSe doping. Our study discovers that the polarization change at the interface Se because of the close proximity to the substrate from that in the free-standing FeSe film significantly amplifies the total potential difference at the interface above and beyond the work function difference for charge transfer. Additionally, the titanate substrate with a large number of free electrons basically serves as an ``infinite'' charge reservoir, which leads to the saturated FeSe doping with the complete removal of interface potential gradient. Our work has developed the theory for modulation doping in the Van der Waals materials/oxides heterostructure, providing a solution to the puzzle of ``magic'' doping in FeSe monolayers on titanates. The information also presents experimental pathways to accommodate a variable carrier density of FeSe monolayers via modulation doping.
\end{abstract}

\maketitle

The bulk iron selenide (FeSe) superconducts below 8\,K \cite{hsu2008superconductivity}, and many methods including applied pressure \cite{medvedev2009electronic}, potassium (K) intercalation \cite{guo2010superconductivity}, and electrical gating \cite{lei2016evolution}, can increase T$_c$ to $\sim$40\,K. A milestone discovery in 2012 is the enhanced superconductivity in a monolayer FeSe on STO  \cite{qiwang-cpl}. Superconducting energy gap opening at $\sim$65\,K is observed in spectroscopy experiments \cite{qiwang-cpl,lee2013interfacial,he2013phase,miyata2015high,tan2013interface,shi2017enhanced}, whereas \textit{in-}\,\&\,\textit{ex-situ} transport measurements exhibit a lower T$_c$$\sim$40\,K \cite{wen2014direct,insitu_4probe,Faeth_transport}. Angle-resolved photoemission spectroscopy (ARPES) studies \cite{lee2013interfacial,he2013phase} reveal that two occupied electron pockets are centered at the 2-Fe Brillouin zone (BZ) corner, \textit{i.e.}, the $M$ point, and the highest occupied states at the $\Gamma$ point, which formed hole pockets in the bulk FeSe, are located $\sim$50\,meV below the Fermi level. The electron doping of FeSe monolayers is manifested by the fractional total charge per unitcell (UC), which also tells that there must be positive charges elsewhere to retain the charge neutrality. The prevailing scenario for the FeSe doping is the transfer of itinerant electrons in substrates to FeSe monolayers, in which the substrate serves as the charge reservoir, analogous to the famous ``modulation doping'' developed for traditional semiconductors. 

\begin{figure*}
	\includegraphics[clip,width=5.4 in]{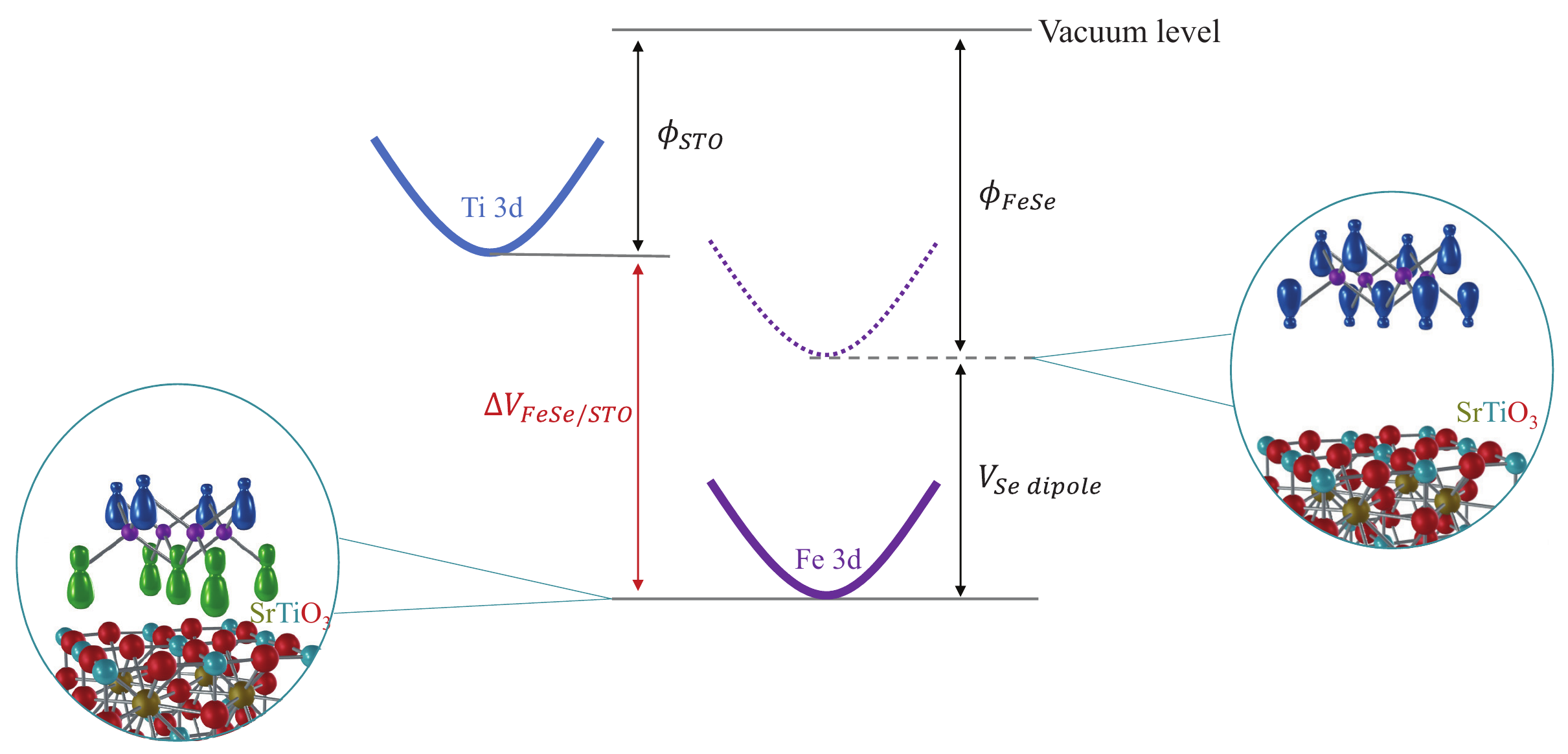}
	\centering
	\caption{The FeSe monolayer and STO band alignment. $\phi_{STO}$ and $\phi_{FeSe}$ denote the work function of STO and FeSe, respectively. $V_{Se\,dipole}$ represents the potential produced by the polarization change at the interface Se. The band offset between Ti 3d and Fe 3d is labeled as $\Delta V_{FeSe/STO}$. The artist's picture shows the Se polarization in the FeSe monolayer on STO (left) and the free-standing film (right).}
\end{figure*}

As demonstrated by the ARPES studies, the low-energy electronic structure of FeSe monolayers on STO \cite{lee2013interfacial,he2013phase} is distinct from that of the bulk systems, such as FeSe with K coating \cite{miyata2015high} or intercalation \cite{K_intercalation}, and (Li,Fe)OHFeSe \cite{zhao2016common}. The point is that, the undoped FeSe monolayer would be predicted to be a semiconductor with an about 10\,meV gap while the bulk FeSe systems without doping are clearly semimetals with the same size of electron and hole pockets. The large difference in the electronic structure suggests that the mechanism for the monolayer superconductivity could be different from that of the bulk one, which is also supported by the higher superconducting gap opening temperature in monolayers \cite{qiwang-cpl,lee2013interfacial,he2013phase,miyata2015high,tan2013interface,shi2017enhanced}. Another impactful finding potentially relevant to the enhanced superconductivity is the replica band observed in ARPES experiments \cite{lee2013interfacial}. The replica ``shake-off'' energy coinciding with a STO phonon energy indicates the remote electron-phonon coupling crossing the interface. This interpretation has been challenged because of the large spacing of $\sim$4\,\AA~between the Fe layer and the STO surface \cite{fangshen_stem,peng2020picoscale} as well as by an alternate possible mechanism for replica bands, namely the energy loss of the emitted photoelectron \cite{fengmiao_eels}. Although no consensus has been reached \cite{klyeshen_replica,liu2021high}, theoretical calculation suggests that the remote electron-phonon coupling would only cause a modest increase in T$_c$ \cite{Dunghai_replica}.

At this point, obtaining the superconductivity phase diagram as a function of electron doping is crucial because it could provide the information for electron pairing itself and the importance of interlayer interactions. In contrast to the dome-like phase diagram of bulk Fe-based superconductors \cite{miyata2015high,stewart_fese_review}, it was reported that the monolayer T$_c$ rises monotonically with the intensity of replica bands and the extended postannealing \cite{he2013phase,song2019evidence}, implying a conventional electron pairing. However, several groups have recently reported that the postannealing process essentially removes Fe vacancies in as-grown films, optimizing the FeSe stoichiometry \cite{huang2016dumbbell,chong_fe_vacancy,xue_stoichiometry}. The K coating can effectively add electrons to FeSe \cite{miyata2015high,tang_k}, but also strongly disturbs the monolayer system due to the electric field generated by the proximate K$^{+}$ ion \cite{choi_k}. All the above urges to develop a reliable monolayer phase diagram with a systematic change of the FeSe carrier density, strongly preferring modulation doping to prevent other possible influences.

However, varying the modulation doping at FeSe/oxide interfaces has been challenging, and an outstanding puzzle from the past decade research is the universal FeSe doping about 0.1 electrons per Fe on a range of Ti oxide substrates \cite{he2013phase,dinghong_sto110,zhang2017ubiquitous,rebec110,peng2014tuning,liu2021_ETO,jia2021magic}. Although other mechanisms for the FeSe doping were suggested \cite{berlijn_2014,jia2021magic}, we will concentrate on the study of charge transfer between the FeSe monolayer and STO using density functional theory (DFT) in order to find the crucial parameter that limits the monolayer doping.


Pure STO is a well-known semiconductor with a $\sim$3.2 eV bandgap, and as mentioned previously, the undoped FeSe monolayer would be a small gap semiconductor. When FeSe and STO are far apart, the separation of Ti 3d and Fe 3d conduction bands simply equalizes the work function difference, as schematically shown in Fig.\,1 and demonstrated by our DFT calculations (Fig.\,S1 in the Supplemental Material \cite{Supplemental}, and see, also references therein \cite{paw,pbe,giannozzi2009quantum,giannozzi2017advanced,unfold,momma2011vesta}.). Here, we define the work function, $\phi_{STO}$ and $\phi_{FeSe}$, as the energy to move electrons from the bottom of Ti and Fe 3d conduction bands to the vacuum level, respectively. The experimental work functions of electron-doped STO with the TiO$_2$ termination $\phi_{STO}$ and that of thick FeSe films $\phi_{FeSe}$ \cite{zhang2017origin,chambers_wk} are equal to $\sim$4.5\,eV and $\sim$5.1\,eV, respectively. Similar to the experiment result, the work function difference between STO and the free-standing FeSe monolayer is about 0.5\,eV in our DFT calculation \cite{Supplemental}.
\begin{figure}[b]
	\includegraphics[clip,width=3.4 in]{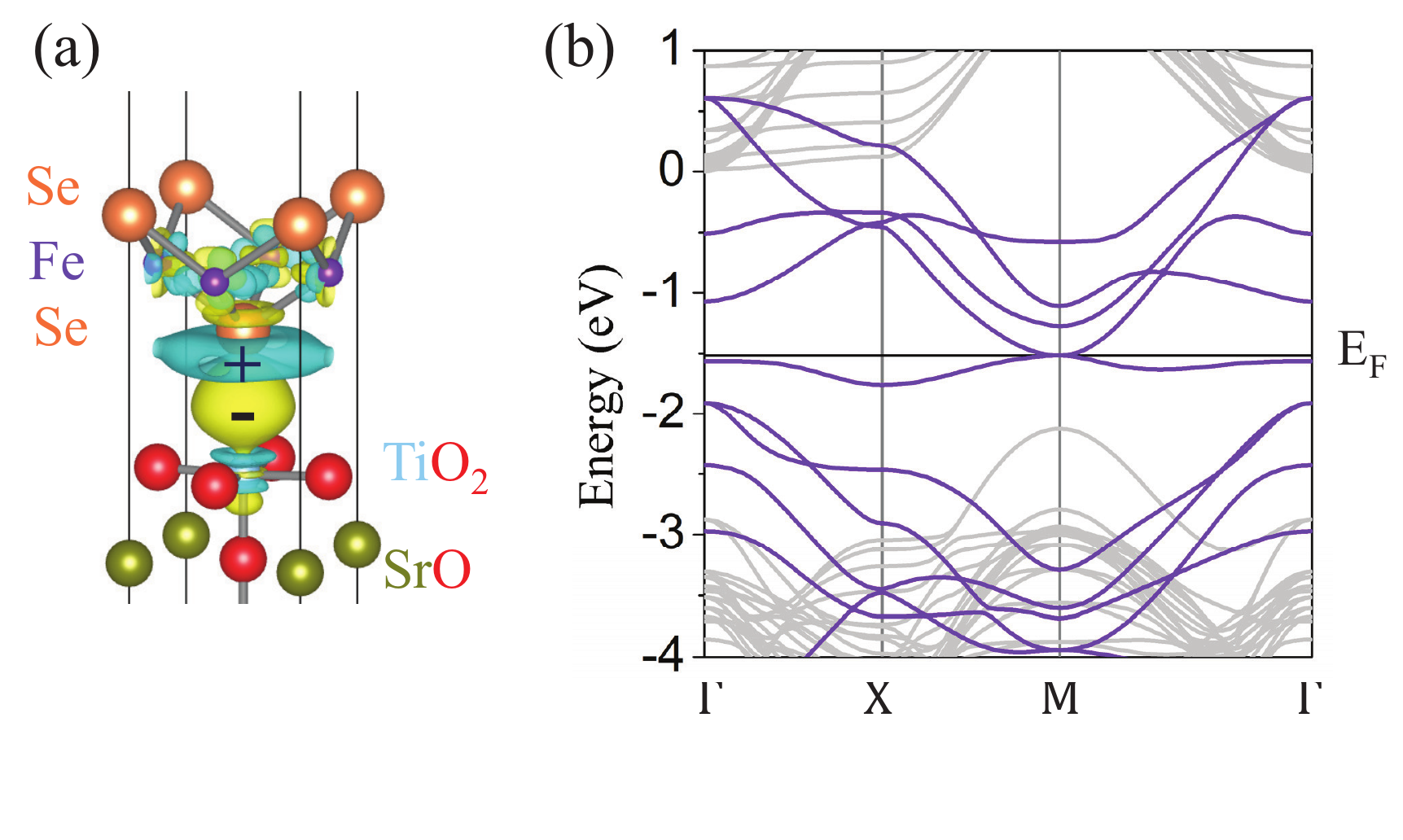}
	\caption{The extra Se dipole at the FeSe/STO interface. (a) the charge density difference calculated using $\rho(FeSe/STO)-\rho(STO)-\rho(FeSe)$ . The isosurface is drawn with the electron density 2.7$\times$10$^{-3}$ e/\AA$^3$. The yellow and cyan colors illustrate the increasing (the net negative charge ``-'') and decreasing (the net positive charge ``+'') electron density, respectively. (b) the DFT-calculated FeSe/STO bandstructure. The purple and gray spaghetti represent the FeSe and STO band dispersion, respectively. The bottom of STO conduction band is set at E=0 as the reference. We adopted the checkerboard antiferromagnetic order in our FeSe monolayer calculations.}	
\end{figure}

\begin{figure*}
	\includegraphics[clip,width=5.2 in]{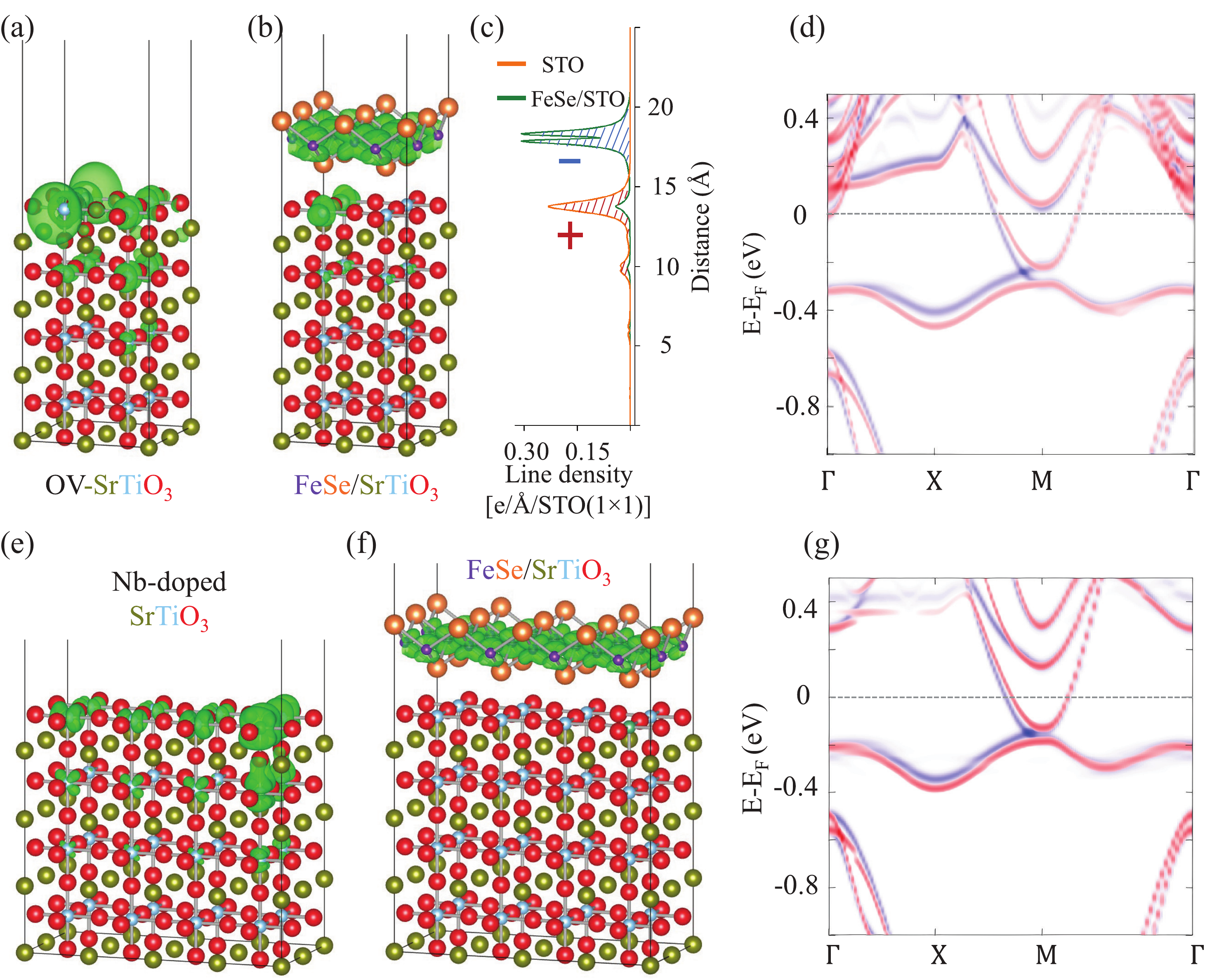}
	\caption{The doping of FeSe monolayers on electron-doped STO. The saturated doping of the FeSe monolayer on the (2$\times$2) STO surface with an OV (a)-(d) and the FeSe doping below the saturation value on the (2$\times$4) STO surface with one Ti replaced by Nb (e)-(g). The green isosurfaces are plotted with the free electron density 3.4$\times$10$^{-3}$ e/\AA$^3$. The red and blue shadow patterns in (c), corresponding to the difference of free electrons in FeSe/STO and STO, illustrate the decrease of electrons (the net positive charge ``+'') on the STO surface and the addition of electrons (the net negative charge ``-'') in the FeSe monolayer. The unfolded bandstructure [(d) and (g)] includes the spin up (red) and down (blue) dispersion.}
\end{figure*}

 In the free-standing FeSe monolayer, both Se at the upper and lower layer are strongly polarized by the middle four-neighboring Fe with the formal valence 2+ \cite{sawatzky2009heavy}. These inherent Se dipoles with large negative lobes pointing toward the central Fe layer, as shown by the artist's picture in Fig.\,1, tend to decrease the net work function of the free-standing film from that value dictated by the minimum electron removal energy if the dipoles were not present. The photoemission experiment to measure the work function of the FeSe film \cite{zhang2017origin} , which removes electrons from the upper Se layer, has taken the effect into account, and the yielded value is close to the work function of the free-standing FeSe film.
 
 Upon the FeSe monolayer approaching the TiO$_2$ surface layer in STO, the change of the whole system mainly happens at the \textbf{interface} Se in close proximity to the tetravalent Ti in the substrate surface, as seen in the calculated charge density difference [Fig.\,2(a)]. The polarization change at the interface Se with a larger electron density on the interface side forms extra dipoles oriented in the direction to effectively increase the potential difference above and beyond the work function difference of the free-stand film and substrate. The calculated bandstructure [Fig. 2(b)] shows the Fe 3d band is shifted downward by an energy equal to the dipole potential $V_{Se\,dipole}$$\sim$1.0\,eV. Consequently, the total potential difference between the two charge neutral systems, \textit{i.e.}, the FeSe monolayer and STO, is given by
 \begin{equation}
 	\Delta V_{FeSe/STO}=V_{Se\,dipole}+(\phi_{FeSe}-\phi_{STO})
 \end{equation}
If the terminating surface in STO is SrO, the size of the extra dipole at the interface Se sitting on the surface Sr$^{2+}$ would be considerably smaller than that on the TiO$_2$ surface (Fig.\,S3 in the Supplemental Material \cite{Supplemental}), revealing the pivotal role of the valence of cations in the substrate surface.

The potential difference between the higher Ti 3d band and the lower Fe 3d conduction band acts as the driving force for the transfer of electrons from STO to FeSe. In order to activate the charge transfer, itinerant electrons in the substrate are required. The ultra-high vacuum (UHV) preannealing at high temperatures to get the smooth substrate surface simultaneously creates a large number of oxygen vacancies (OVs) in STO \cite{chen2015observation}. Also, the chemical substitution such as Nb$^{5+}$ (La$^{3+}$) replacing Ti$^{4+}$ (Sr$^{2+}$) \cite{tomioka2019enhanced} is commonly used to obtain the conducting STO. As seen in Fig.\,3(a) \& (e), the free electrons in STO are disproportionately distributed on the surface because of the lower Ti 3d surface state \cite{Supplemental}, forming the well-known two-dimensional electron gas (2DEG) \cite{meevasana2011creation,santander2011two}.

Once FeSe is grown on electron-doped STO, the free electrons in the substrate will migrate to the FeSe monolayer in order to lower the whole system's total energy. The electron transfer is seen in our DFT calculations of FeSe monolayers on an oxygen-vacant and Nb-doped STO substrates (Fig.\,3 and S5 in the Supplemental Material \cite{Supplemental}), in consistence with previous DFT studies \cite{bang_2013,chen_2016}. Structurally, the electron-doped FeSe monolayer tends to move closer to the substrate surface with the net positive charge. In addition, we have observed clear difference in spin-up and -down band dispersion of electron-doped FeSe monolayers close to the Fermi level [Fig.\,3(d) \& (g)], which  could be possibly due to the interaction between the electric field in between and the checkerboard-type magnetic order.

The electron redistribution at the interface, in essence, forms a ``capacitor''-like structure consisting of the electron-doped Fe plane and the Ti oxide surface layer with the majority of net positive charges [Fig.\,3(c)]. This creates the built-in potential in between approximately given by V$_{bi}$=$\frac{\sigma}{\varepsilon}$d, where $\sigma$, $d$, and $\varepsilon$ denote the charge density and the distance of two charged planes, and the dielectric constant of the material in between, respectively. Assuming $\varepsilon=1$ and using other parameters from experiments, we yield $V_{bi}$$\sim$1.0\,eV, which is close to the calculated potential difference $\Delta V_{FeSe/STO}$. According to the capacitor model, the energy gain of the whole system due to charge transfer is proportional to $\sim$$\frac{1}{2}$$\sigma$V$_{bi}$.

The number of transferred electrons is limited by the capacitor built-in potential when it exactly equalizes $\Delta V_{FeSe/STO}$. However, this level of the FeSe doping can only be reached in the condition that the density of free electrons in the substrate $n_{sub}$ is no less than that required to completely remove the potential difference at the interface, $\sigma_0$. As shown in Fig.\,3(b) and Fig.\,5S, the majority of free electrons in STO is transferred to the FeSe monolayer while some electrons still stay in the substrate, which indicates that the FeSe doping is saturated. The Fermi level crossing the STO conduction band demonstrates the complete removal of the potential difference $\Delta V_{FeSe/STO}$, confirming that the FeSe electron density is pinned at the saturated level compensating for the potential difference. The larger electron pocket in our DFT calculation [Fig.\,3(d)] compared to that in experiments is mainly because only one electron band at M is present while the actual system has two bands \cite{doublebands}. We note that this is not unexpected in a system in which there is no long-range magnetic order but strong short-range correlations of the same kind. 

When $n_{sub}<\sigma_0$, the charge transfer would be out-of-equilibrium. In this scenario, all itinerant electrons in substrates would be transferred to the FeSe monolayer [Fig.\,3(e)-(g)], which, however, creates a smaller electron pocket compared with that in Fig.\,3(d) because it is $n_{sub}$ that limits the FeSe doping below the saturation value. The total potential difference at the interface is not eliminated by the charge transfer. In Fig.\,3(g), the energy difference between the Fermi level and the bottom of the STO conduction band corresponds to the uncompensated potential difference.

In light of the very small number of electrons demanded by the FeSe monolayer, namely $\sigma_0$$\sim$\,0.1 electrons per Fe, the STO substrate preannealed in UHV or doped with Nb (La) with a large number of free electrons basically serves as an ``infinite'' charge reservoir. This also applies to other reported FeSe/Ti oxide interfaces \cite{dinghong_sto110,zhang2017ubiquitous,rebec110,peng2014tuning,liu2021_ETO,jia2021magic} where the equilibrium charge transfer always occurs. The tetravalent Ti in these substrate surfaces induces similar polarization changes at the interface Se, which, together with the similar experimental work functions \cite{chambers_wk,mansfeldova2021work,gerholdsto110}, results in nearly equal potential differences. It in turn leads to almost substrate independent FeSe doping, 0.1$\pm$0.02 electrons per Fe. The about 20\% doping variation \cite{dinghong_sto110,zhang2017ubiquitous,rebec110,peng2014tuning,liu2021_ETO,jia2021magic} is attributed to some differences in the work function and the details of the Ti oxide surface layers resulting from different crystal orientations and atomic structures of these titanates. Recently, the universal doping is expanded to the FeSe monolayer on Fe oxide substrates which have the work function $\sim$4.6\,eV \cite{song_lfo}. This suggests that the polarization at the interface Se of the FeSe monolayer sitting on top of the Fe in the substrate surface has a similar contribution to the interface potential difference compared with that on the Ti oxide substrates.

In order to change the saturation value of FeSe doping, one can certainly try other electron-doped substrates with very different work functions or distinct valences of surface cations resulting in changes in the interface Se polarization. Also, the interface potential could be possibly modulated by electrical gating. To vary the FeSe carrier density below the saturation value, One could, in principle, use substrates with electron densities lower than that required to reach the equilibrium value of the potential difference at the interface. 

In summary, we have developed the theory for modulation doping in FeSe monolayers on STO, and found that: (1) the presence of the extra dipole at the interface Se due to the close proximity to the substrate strongly enhances the charge transfer; (2) the doping of FeSe monolayers on a range of Ti oxide substrates studied to date is saturated at a similar level because of the effectively ``infinite'' number of free electrons in substrates and the similarities of the substrates which all have formally 4+ Ti cations in the terminating surface. The theory also applies to the charge redistribution at other interfaces between oxides and two-dimensional Van der Waals materials, where the highly polarizable anions in the interface layer of the Van der Waals material are in close proximity to oxide cations at the interface. Our studies have identified the crucial parameter limiting modulation doping of FeSe monolayers on STO, paving the way for the development of a reliable superconducting phase diagram with a possibly higher monolayer T$_c$ at an, as yet unknown, doping level.

\begin{acknowledgments}
We thank C. Liu and K. Zou for discussions. The work was supported by Natural Sciences and Engineering Research Council of Canada (NSERC) and Canada Foundation for Innovation (CFI). This research was undertaken thanks in part to funding from the Max Planck-UBC-UTokyo Centre for Quantum Materials and the Canada First Research Excellence Fund, Quantum Materials and Future Technologies Program.
\end{acknowledgments}
 
\bibliography{ref}
 
\end{document}